\documentclass[journal,twocolumn,letterpaper]{IEEEJERM}

\ifCLASSINFOpdf
 
\else

\fi

\usepackage{times,amsmath,epsfig}
\renewcommand\IEEEkeywordsname{Index Terms}
\usepackage{fancyhdr}
\usepackage{amsmath}
\usepackage{amsfonts}
\usepackage{amssymb}
\usepackage[utf8]{inputenc}
\usepackage{array}
\usepackage{graphicx}
\usepackage{url}
\usepackage{subfigure}
\usepackage{bm}
\usepackage{breqn}
\usepackage{xcolor}
\usepackage{soul}
\usepackage{amssymb}
\usepackage{flushend}
\usepackage{graphicx}
\usepackage{multirow}
\usepackage{booktabs}

\IEEEoverridecommandlockouts

\begin{document}

\title{GRN-Transformer: Enhancing Transformer's Performance for Artifact Detection in PICU Photoplethysmogram Signals}

\author{Thanh-Dung Le,~\IEEEmembership{Senior Member,~IEEE,} Clara Macabiau,  Kevin Albert, \\
    Philippe Jouvet, and Rita Noumeir,~\IEEEmembership{Member,~IEEE}

 \thanks{This work was supported in part by the Natural Sciences and Engineering Research Council (NSERC), in part by the Institut de Valorisation des données de l’Université de Montréal (IVADO), in part by the Fonds de la recherche en sante du Quebec (FRQS).}

\thanks{Thanh-Dung Le is with the Biomedical Information Processing Lab, \'{E}cole de Technologie Sup\'{e}rieure, University of Qu\'{e}bec,  Montr\'{e}al, Qu\'{e}bec, Canada, and is with The Interdisciplinary Centre for Security, Reliability, and Trust, University of Luxembourg, Luxembourg (Email: thanh-dung.le@uni.lu).}

\thanks{Clara Macabiau is with the Biomedical Information Processing Lab, \'{E}cole de Technologie Sup\'{e}rieure, University of Qu\'{e}bec,  Montr\'{e}al, Qu\'{e}bec, Canada.}

\thanks{Kevin Albert is with the CHU Sainte-Justine Research Center, CHU Sainte-Justine Hospital, University of Montreal, Montr\'{e}al, Qu\'{e}bec, Canada.}

\thanks{Philippe Jouvet is with the CHU Sainte-Justine Research Center, CHU Sainte-Justine Hospital, University of Montreal, Montr\'{e}al, Qu\'{e}bec, Canada.}

\thanks{Rita Noumeir is with the Biomedical Information Processing Lab, \'{E}cole de Technologie Sup\'{e}rieure, University of Qu\'{e}bec,  Montr\'{e}al, Qu\'{e}bec, Canada.}
 
}

\markboth{IEEE }
{T.D. Le \MakeLowercase{\textit{et al.}}: }

\twocolumn[
\begin{@twocolumnfalse}
  
\maketitle

\begin{abstract}

Photoplethysmogram (PPG) signals, optical measurements of pulsatile blood flow used continuously in intensive care monitoring, are frequently contaminated by motion, low perfusion, or sensor displacement, producing waveform artifacts that propagate into downstream estimates such as SpO\textsubscript{2} and trigger spurious clinical alarms. Automated artifact detection is therefore a prerequisite for reliable bedside decision support. Transformer classifiers, which use self-attention to weight contributions from every part of the input pulse, are well suited to learning artifact morphology, but their performance is known to degrade on the small, imbalanced datasets typical of single-center clinical studies. We propose the \emph{GRN-Transformer}, which integrates a single Gated Residual Network (GRN) block atop a standard Transformer encoder stack, serving as a small-data regularizer. On a labeled Pediatric Intensive Care Unit (PICU) PPG dataset from CHU Sainte-Justine Hospital (CHUSJ), the GRN-Transformer reaches $98\%$ accuracy, $90\%$ precision, $97\%$ recall, and $93\%$ F1-score using only $5\%$ of the annotated pulses, substantially improving recall over the baseline Transformer ($+11$ points) without sacrificing precision. Cross-validated evaluation indicates that clean-data accuracy gains are modest once fold-to-fold variability is accounted for, but the GRN-Transformer is markedly more robust to realistic signal degradations (noise, baseline wander, sensor dropout), trains approximately $2.7\times$ faster than the baseline, and runs at $6.33$~ms p99 latency on a CPU-only consumer laptop. A retrospective simulation suggests the model could meaningfully reduce clinician review burden when used as a pre-filter for PPG-driven alarms. These results support the GRN-Transformer as a deployable artifact-detection component for resource-constrained pediatric clinical settings. 

\vspace{9pt}
\begin{IEEEkeywords}
clinical PPG signals, Transformers, Gated Residual Networks, imbalanced classes, and artifact detection.
\end{IEEEkeywords}

\vspace{9pt}

\textbf{\textit{Clinical and Translational Impact Statement---} Reliable PPG artifact detection is a prerequisite for clinical decision support in pediatric intensive care: undetected artifacts propagate into erroneous SpO\textsubscript{2} estimates and trigger spurious desaturation alarms. On PPG data from the PICU at CHUSJ, the proposed GRN-Transformer raises recall from $86\%$ to $97\%$ over a standard Transformer, runs at $6.33$~ms p99 latency on a CPU-only laptop, and degrades gracefully under realistic signal corruption. A retrospective simulation indicates the model could substantially reduce artifact-induced alarm burden, supporting translational use in PICU monitoring workflows pending prospective external validation.}
\end{abstract}

\end{@twocolumnfalse}]

{
  \renewcommand{\thefootnote}{}%
  \footnotetext[1]{This work was supported in part by the Natural Sciences and Engineering Research Council (NSERC), in part by the Institut de Valorisation des donnees de l'Universite de Montreal (IVADO), in part by the Fonds de la recherche en sante du Quebec (FRQS), and in part by the Fonds de recherche du Quebec-Nature et technologies (FRQNT). }

  \footnotetext[2]{Thanh-Dung Le, Clara Macabiau, and Rita Noumeir are with the Biomedical Information Processing Lab, Ecole de Technologie Superieure, University of Quebec, Canada (Email: thanh-dung.le@tamucc.edu).}

  \footnotetext[3] { Kevin Albert and Philippe Jouvet are with the Research Center at CHU Sainte-Justine, University of Montreal, Canada.}
}
 
\IEEEpeerreviewmaketitle

\section{Introduction}
\IEEEPARstart{P}{ICU}s increasingly rely on clinical decision support systems (CDSS) that draw on continuously recorded physiological signals~\cite{dziorny2022clinical}. A central input to such systems is peripheral oxygen saturation (SpO\textsubscript{2})~\cite{le2022detecting, sauthier2021estimated}, which is estimated from the photoplethysmogram (PPG), an optical measurement of pulsatile blood flow obtained from a fingertip or earlobe pulse oximeter. PPG signals are easily corrupted by patient movement, sensor displacement, and low peripheral perfusion; the resulting waveform artifacts produce erroneous SpO\textsubscript{2} estimates that propagate into spurious desaturation alarms and degrade the reliability of any CDSS that consumes them. Automated artifact detection at the input stage is therefore a prerequisite for safe PPG-driven decision support, particularly in the PICU setting at CHUSJ, where a high-resolution research database links bedside biomedical signals to electronic patient records~\cite{mathieu2021validation, dziorny2022clinical}. Our recent work~\cite{claramacabiau2023} examined several machine learning (ML) approaches for PPG artifact detection in this setting and found that, under the small-data, class-imbalanced regime typical of single-center clinical studies, simpler methods such as $ K$-nearest-neighbor classification and semi-supervised label propagation outperformed Transformer models. Transformers, which use a self-attention mechanism to let each part of the input pulse learn to weight the contributions of every other part, capturing long-range morphological dependencies without recurrence, have a strong representational prior for this task, but their effectiveness on small datasets remains an open challenge~\cite{lee2021vision, shao2022transformers}: scaling parameters up to compensate tends to produce overfitting in data-constrained regimes. This study addresses that challenge by integrating a GRN block into the Transformer classifier. The resulting \emph{GRN-Transformer} retains the encoder backbone of the standard Transformer but adds a single gated residual block on the pooled output, acting as a small-data regularizer rather than an attention-mechanism modification. We characterize the architecture along four axes that the original PICU-PPG benchmark did not address: single-split and cross-validated classification quality; interpretability of the gated representation; robustness to realistic PPG signal degradations; and deployment characteristics, including inference latency on a CPU and a retrospective estimate of clinician review burden. Taken together, these analyses position the GRN-Transformer as a deployment-ready artifact-detection component for resource-constrained pediatric clinical settings.

\section{Related Work} ML approaches to PPG signal analysis span a broad range of clinical applications including heart-rate estimation~\cite{dao2016robust, mehrgardt2021deep}, blood-pressure estimation~\cite{liu2020pca}, respiratory-rate extraction~\cite{birrenkott2017robust}, and quality assessment~\cite{venema2013robustness, alharbi2022non, nwibor2023remote}. For artifact detection specifically, deep models such as multilayer perceptrons (MLPs) and fully convolutional networks (FCNNs) have largely supplanted classical classifiers~\cite{wang2017time, marzorati2022hybrid}, and time-domain features, combined with bidirectional LSTM models, have produced strong results in heart-rate-related quality assessment~\cite{maqsood2021benchmark}. Within this landscape, our prior work on PICU PPG artifact detection~\cite{claramacabiau2023} found that under the small-data, class-imbalanced regime typical of single-center PICU studies, simpler approaches, semi-supervised label propagation and $K$-nearest-neighbor classification, outperformed both MLP and Transformer baselines. Transformers nevertheless retain a strong representational prior for sequence classification, owing to the self-attention mechanism's capacity to model long-range dependencies between any two positions in an input pulse~\cite{lin2022survey}. The challenge is to recover that representational advantage without simply scaling the model up, which often leads to overfitting in data-constrained regimes~\cite{claramacabiau2023, mehrgardt2021deep}. Existing remedies modify the attention mechanism, add data augmentation~\cite{lee2021vision}, or hybridize attention with convolutional layers~\cite{shao2022transformers}; the latter inherits the quadratic-in-sequence-length cost of self-attention and adds the sequential-locality bias of convolutions, both of which limit applicability on long physiological windows~\cite{hahn2020theoretical, sattler2019understanding}. The present work takes an alternative route: rather than modifying the attention mechanism itself, we add a single gated residual block atop the encoder stack as a small-data regularizer. The next subsection details the GRN design and its placement in the architecture.

\begin{figure*}[t] \centering \includegraphics[width=\textwidth]{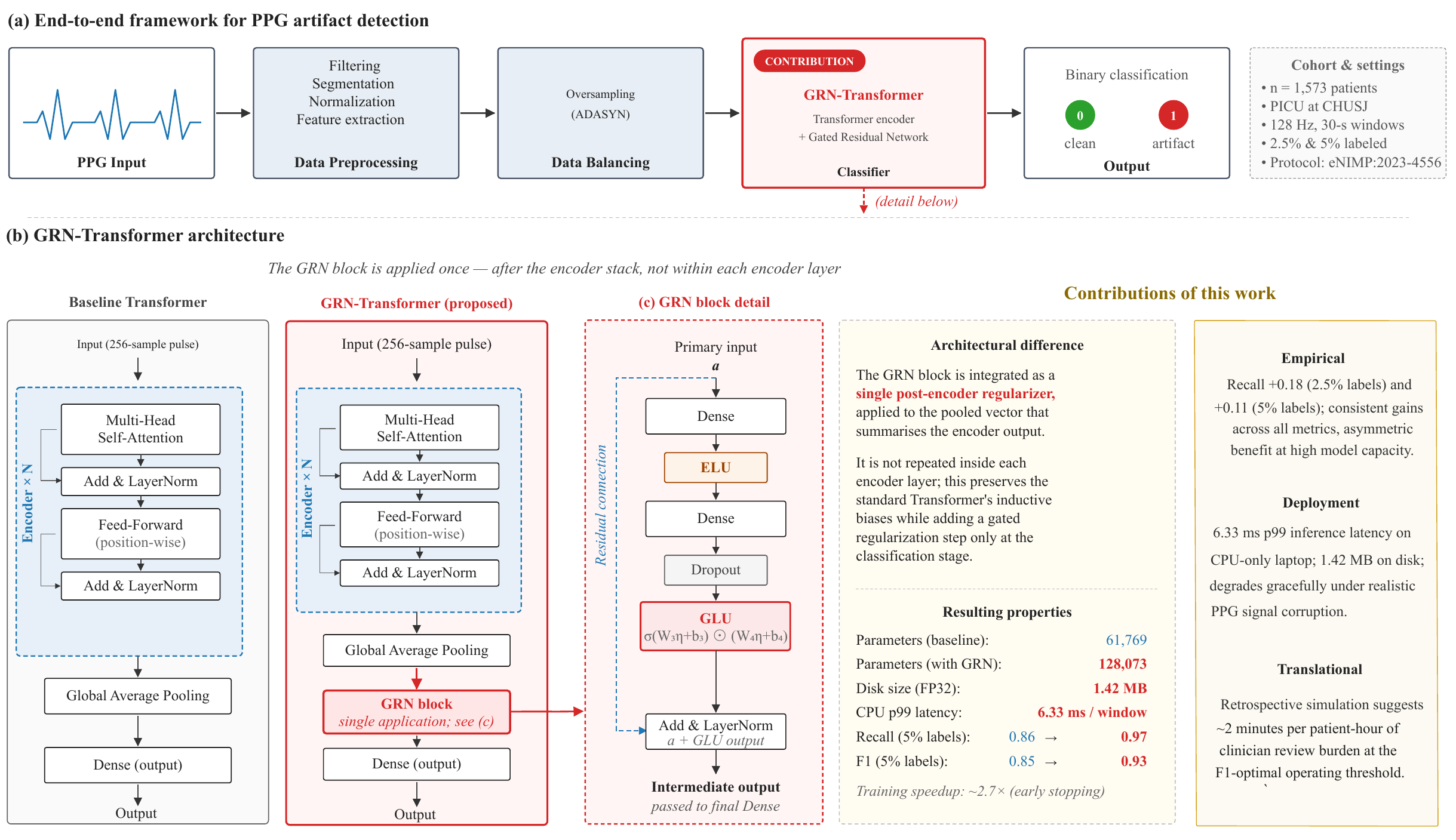} 
\caption{The proposed GRN-Transformer framework. (a)~End-to-end pipeline for PPG artifact detection at the CHU Sainte-Justine PICU: raw PPG signals are preprocessed (filtering, segmentation, resampling, feature extraction), balanced via ADASYN oversampling, classified by the GRN-Transformer, and labeled as clean (0) or artifact (1). (b)~Architecture comparison between the baseline Transformer and the proposed GRN-Transformer. The GRN block is integrated as a \emph{single} post-encoder regularizer, applied once to the pooled output of the $N$-layer encoder stack rather than repeated within each encoder layer; this preserves the standard Transformer's inductive biases while adding gated regularization at the classification stage. The side panel summarises the architectural difference and reports the resulting parameter count, disk footprint, CPU inference latency, and performance gains at the $5\%$-labeled regime. (c)~Internal structure of the GRN block: a $\mathrm{Dense}\!\rightarrow\!\mathrm{ELU}\!\rightarrow\!\mathrm{Dense}$ nonlinear branch, dropout, and a Gated Linear Unit (GLU), combined with the residual identity connection through an add-and-normalize layer (Eqs.~(1)--(5)).} \label{fig:framework} 
\end{figure*}

\section{Materials and Methods}

Fig.~\ref{fig:framework} summarises the end-to-end framework employed in this study. Panel~(a) shows the four-stage pipeline: acquisition of PPG signals from the PICU at CHUSJ (subject to the inclusion and exclusion criteria detailed in Section~\ref{sec:data_CHUSJ}); preprocessing for filtering, segmentation, resampling, and feature extraction; class-imbalance correction via ADASYN oversampling; and binary classification of each $30$-second window as artifact or clean. The GRN-Transformer classifier sits at the centre of this pipeline; its architectural detail is shown in panels~(b) and~(c) and described in Section~\ref{sec:grn}.

\subsection{Clinical PPG Data at CHUSJ} \label{sec:data_CHUSJ} The study was approved by the research ethics board of CHUSJ (project number eNIMP:2023-4556). The study population comprised all children aged $0$ to $18$ years who were admitted to the PICU between September 2018 and September 2023 and had available PPG waveform records. Data collected beyond the fourth day of hospitalization were omitted to avoid bias from patients with prolonged arterial-line stays, and patients undergoing extracorporeal membrane oxygenation were excluded. For patients with multiple admissions, only the first stay was retained to ensure data independence. After applying these criteria, $1{,}573$ patients were included. For each patient, PPG signals were acquired at a sampling frequency of $128$~Hz in $5$-second windows, continuously across the first $96$~hours of admission. A fixed $30$-second window was used as the analysis unit for all subsequent processing.

\subsection{Data Preprocessing}
\label{sec:data_prep}

\begin{figure}[!htp]
	\centering
	\includegraphics[scale=0.65]{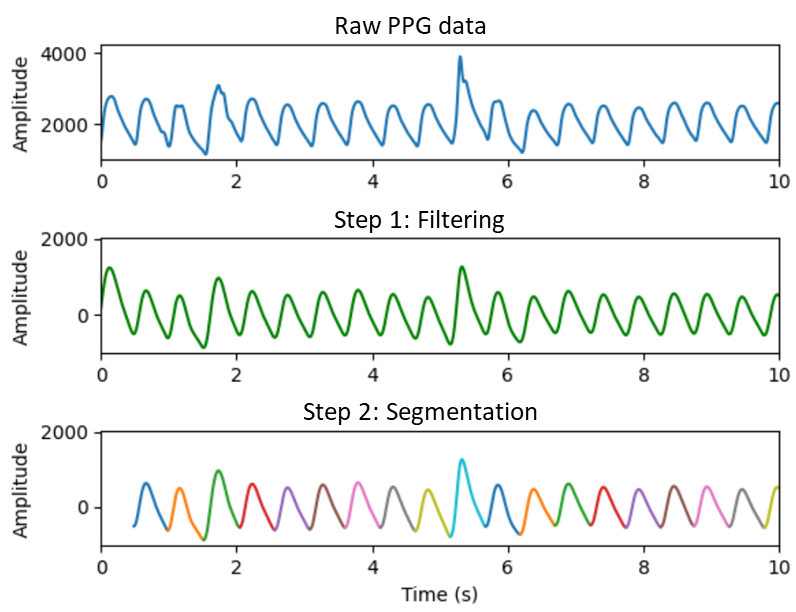}
    \caption{Illustration of the first two preprocessing stages on a representative $10$-second window. \emph{Top:} raw PPG signal as recorded by the bedside pulse oximeter, with non-zero baseline, slow baseline drift, and a visible amplitude excursion near $t \approx 5.3$~s. \emph{Middle:} the same window after Butterworth bandpass filtering ($0.5$--$5$~Hz, forward--backward), which removes baseline drift and centers the signal at zero while preserving pulse morphology. \emph{Bottom:} pulse-level segmentation by consecutive local minima; each color marks one pulse, the basic unit of classification. The anomalous pulse at $t \approx 5.3$~s is now isolated into a single segment that visibly departs from the morphology of its neighbors, exemplifying the kind of motion artifact the downstream classifier must identify. Procedure reproduced from~\cite{claramacabiau2023}.}
	\label{fig:preprocessing}
\end{figure}

\begin{table}[!tp]
\centering
\caption{Statistical summary of the dataset}
\begin{tabular}{|l|c|c|c|}
\hline
Statistic                & Overall & Non-artifact & Artifact\\ \hline
Count                             & 8190             & 6753             & 1437             \\ \hline
Mean                              & 13.53            & 14.98            & 6.70             \\ \hline
Standard Deviation                & 329.36           & 285.95           & 439.81           \\ \hline
Minimum                           & -1784.64         & -1590.12         & -1686.26         \\ \hline
25th Percentile                   & -185.99          & -165.09          & -254.19          \\ \hline
50th Percentile (Median)          & -5.05            & -1.61            & -12.73           \\ \hline
75th Percentile                   & 203.48           & 180.45           & 279.35           \\ \hline
Maximum                           & 2016.81          & 1644.23          & 1981.25          \\ \hline
Skewness                          & 0.23             & 0.39             & -0.00            \\ \hline
Kurtosis                          & 3.04             & 3.28             & 1.70             \\ \hline
\end{tabular}
\label{table:summary_statistics}
\end{table}

Table~\ref{table:summary_statistics} reports descriptive statistics
of the dataset, separated by class. The two classes have similar
central tendencies; both medians sit near zero after
normalization, but differ substantially in their distributional
shape. Artifact pulses exhibit roughly $1.5\times$ the standard
deviation of clean pulses ($439.8$ vs $286.0$) and a lower kurtosis
($1.70$ vs $3.28$), indicating a wider, flatter amplitude
distribution. Clean pulses cluster tightly around a physiological
mode, whereas artifacts are spread more uniformly across the full
amplitude range. These distributional differences are the discriminative signal that the downstream classifiers must learn.

Each PPG recording was processed in three steps: filtering, segmentation, and resampling. \textbf{Step 1 (Filtering):} Each $30$-second window was bandpass filtered using a Butterworth filter with cutoff frequencies at $0.5$~Hz and $5$~Hz, corresponding to a $30$--$300$~bpm heart-rate range. Forward--backward filtering was applied to eliminate phase distortions, removing baseline wander and high-frequency noise. \textbf{Step 2 (Segmentation):} Pulse boundaries were identified through local-minima detection on the filtered signal. Each pulse was defined as the segment between two consecutive minima, enabling assessment of artifacts at the level of an individual heartbeat. \textbf{Step 3 (Resampling and Normalization):} Each pulse was uniformly resampled to $256$ samples per beat ($4$~ms per sample, equivalent to a one-second cardiac cycle), using linear interpolation to estimate missing points~\cite{li2012dynamic}. The resampled pulses were normalized to zero mean and unit variance to ensure consistent scaling across input features. Ground-truth labels were established by a healthcare professional who manually annotated each PPG pulse as artifact or normal based on its morphology. To verify annotation consistency, an automated re-annotation algorithm previously developed by our group~\cite{claramacabiau2023} re-examined $10\%$ of the expert-annotated pulses by applying statistical checks against expected pulse parameters; disagreements were resolved by re-inspection. To assess how much labeled data is required for reliable artifact detection, we evaluated the classifiers on two annotated subsets: $2.5\%$ ($n=2{,}137$ pulses) and $5\%$ ($n=4{,}170$ pulses) of the full database. Across both subsets, the class distribution is moderately imbalanced (Table~\ref{tab:data_imbalance}), with $17$--$18\%$ artifact pulses in line with the prevalence reported in related PPG quality assessment studies~\cite{claramacabiau2023}. Class imbalance was handled at training time using the ADASYN oversampling method~\cite{he2008adasyn}.

\begin{table}[!t]
\footnotesize
\caption{Annotated data proportion and its imbalanced characteristics}
\label{tab:data_imbalance}
\begin{tabular}{|l|l|l|l|l|}
\hline
Data Portion & Artifacts (\%) & Normal (\%) & Dimension          & Level \\ \hline
2.50\%          & 17.3           & 82.7        & 2137x256        & Moderate            \\ \hline
5\%             & 18.1           & 81.9        & 4170x256             & Moderate            \\ \hline
\end{tabular}
\end{table}

\subsection{Machine Learning Classifiers}
\label{sec:ml_classifiers}

Several studies have explored the use of deep learning algorithms, such as MLPs and FCNs, for artifact detection, yielding promising results \cite{wang2017time, marzorati2022hybrid}. Recent research has emphasized the effectiveness of time-domain features in conjunction with deep learning algorithms for artifact detection in PPG signals \cite{maqsood2021benchmark}, and the BiLSTM model incorporating time-domain features has demonstrated superior performance for heart rate estimation compared to other models across multiple datasets. Additionally, our research team's investigation \cite{claramacabiau2023, le2024novel} has verified the feasibility of employing diverse ML techniques, including semi-supervised learning label propagation, conventional ML, MLP, and Transformer, for PPG artifact detection. Given these findings, this study concentrates on these benchmarks and baselines for our classifiers. Specifically, we will focus on MLP, FCNN, BiLSTM, and Transformer classifiers.

\subsection{Gated Residual Networks}
\label{sec:grn}
To address the small-data challenge identified above, we incorporate a GRN as a structural component of the Transformer-based classifier. The residual gate is designed to learn identity mappings rapidly, allowing the optimizer to propagate information through the network with minimal interference and to engage the nonlinear branch only when it improves the representation~\cite{savarese2017residual, yang2020interpolation}. Architecturally, this provides a regularizing inductive bias that is particularly effective in data-constrained regimes~\cite{lim2021temporal}, distinguishing the GRN from alternative remedies such as attention-mechanism modifications~\cite{lee2021vision} or attention-CNN hybrids~\cite{shao2022transformers}. The central component of the GRN is the Gated Linear Unit (GLU) \cite{dauphin2017language}, which enables the network to suppress or amplify information based on the task. Following the formulation of Temporal Fusion Transformers~\cite{lim2021temporal}, the GRN takes a primary input $a$ and produces: 

\begin{align} 
\text{GRN}_\omega(a) &= \text{LayerNorm}(a + \text{GLU}_\omega(\theta_1)), \\ 
\theta_1 &= W_{1,\omega}\,\theta_2 + b_{1,\omega},  \\ 
\theta_2 &= \text{ELU}(W_{2,\omega}\,a + b_{2,\omega}) \end{align} 

where $\theta_1, \theta_2$ are intermediate representations, $\text{LayerNorm}$ denotes standard layer normalization~\cite{ba2016layer}, $\omega$ indicates weight sharing, and $\text{ELU}(\cdot)$ is the Exponential Linear Unit: 
\begin{align} 
\text{ELU}(x) = \begin{cases} x & \text{if } x > 0 \\
\alpha(\exp(x) - 1) & \text{if } x \leq 0 \end{cases} 
\end{align} 

with $\alpha > 0$. The GLU itself, given an input $\eta$, is defined as: 
\begin{align} 
\text{GLU}_\omega(\eta) = \sigma(W_{3,\omega}\,\eta + b_{3,\omega}) \odot (W_{4,\omega}\,\eta + b_{4,\omega}), \label{eqn:component_gate} 
\end{align} 

where $W_{(\cdot)}, b_{(\cdot)}$ are weights and biases, $\odot$ is the element-wise Hadamard product, and $\sigma(\cdot)$ is the sigmoid activation. The sigmoid output controls the contribution of the GRN's nonlinear branch to the residual sum: when the GLU output approaches zero, the input $a$ is forwarded essentially unchanged through the residual connection; when it approaches one, the nonlinear branch fully contributes. During training, dropout is applied to $\theta_1$ before the gating layer, and layer normalization is applied to improve robustness and prevent overfitting~\cite{savarese2017residual}.


The GRN block is integrated into the Transformer classifier from our prior study~\cite{claramacabiau2023} as a \emph{single} post-encoder component, applied once to the pooled output of the Transformer encoder stack rather than repeated within each encoder layer. Fig.~\ref{fig:framework}(b) shows the integration alongside the baseline Transformer for comparison: the encoder stack is unchanged, a global average pooling step collapses the temporal dimension to a fixed $d$-dimensional vector, the GRN block (detailed in Fig.~\ref{fig:framework}(c)) operates on this pooled vector through the gated identity transformation of Eqs.~(1)--(5), and a final dense layer produces the binary artifact/clean prediction. This placement preserves the inductive biases of the standard Transformer encoder while adding a single gated-residual regularization step at the classification stage; it also explains the architecture-specific pattern of empirical gains reported in Section~\ref{sec:overall-comparison}, where applying the same GRN block to the pooled outputs of the MLP, FCNN, and BiLSTM baselines produces negligible improvement.

\begin{table*}[]
\centering
\caption{Hyperparameters of classifiers}
\label{tab:my_hyperparameters}
\begin{tabular}{|l|l|l|l|l|l|}
\hline
Hyperparameters                   & Transformer & LSTM     & FCNN     & MLP      & GRN  \\ \hline
Hidden layers                     & 4           & 2        & 3        & 3        & 2    \\ \hline
Number of   neurons               & 128         & 500      & 64       & 500      & 128  \\ \hline
Number of   multi-heads attention & 4           & N/A      & N/A      & N/A      & N/A  \\ \hline
Batch size                        & 96          & 96       & 96       & 96       & N/A  \\ \hline
Dropout                           & 0.25        & 0.3      & 0.25     & 0.3      & 0.25 \\ \hline
Learning rate                     & 6e-04       & 1e-04     & 1e-04     & 1e-04 & N/A  \\ \hline
Optimizer                         & Adam        & Adam     & Adam     & Adam     & N/A  \\ \hline
\end{tabular}
\end{table*}

\section{Experimental Results} 
\label{sec:results}

\begin{table*}[!htp]
\centering
\footnotesize
\caption{Overall performance comparison of all classifiers with and without Gated Residual Networks, on the 2.5\% and 5\% labeled subsets. Best F1 per data block in bold; overall best in italic-bold.}
\label{tab:my-table}
\begin{tabular}{llccccccccc}
\toprule
& & \multicolumn{4}{c}{2.5\% Data} & & \multicolumn{4}{c}{5\% Data} \\
\cmidrule(lr){3-6} \cmidrule(lr){8-11}
& Model & Acc & Pre & Rec & F1 & & Acc & Pre & Rec & F1 \\
\midrule
\multirow{4}{*}{W/o GRN}
  & MLP         & \textbf{0.96} & 0.86 & 0.94 & \textbf{0.90} & & 0.96 & 0.89 & 0.89 & 0.89 \\
  & FCNN        & 0.95 & 0.84 & 0.86 & 0.85 & & 0.95 & 0.86 & 0.83 & 0.84 \\
  & BiLSTM      & \textbf{0.96} & 0.85 & \textbf{0.96} & \textbf{0.90} & & 0.97 & \textbf{0.91} & 0.94 & 0.92 \\
  & Transformer & 0.94 & 0.86 & 0.77 & 0.81 & & 0.95 & 0.85 & 0.86 & 0.85 \\
\midrule
\multirow{4}{*}{With GRN}
  & MLP         & \textbf{0.96} & 0.88 & 0.93 & \textbf{0.90} & & 0.96 & 0.86 & 0.94 & 0.89 \\
  & FCNN        & 0.95 & 0.85 & 0.84 & 0.85 & & 0.94 & 0.83 & 0.80 & 0.81 \\
  & BiLSTM      & \textbf{0.96} & 0.88 & 0.93 & \textbf{0.90} & & 0.97 & 0.92 & 0.89 & 0.90 \\
  & Transformer & \textbf{0.96} & \textbf{0.87} & 0.95 & \textbf{0.90} & & \textit{\textbf{0.98}} & 0.90 & \textit{\textbf{0.97}} & \textit{\textbf{0.93}} \\
\bottomrule
\end{tabular}
\end{table*}

\begin{figure*}[t] 
\centering 
\includegraphics[width=0.95\textwidth]{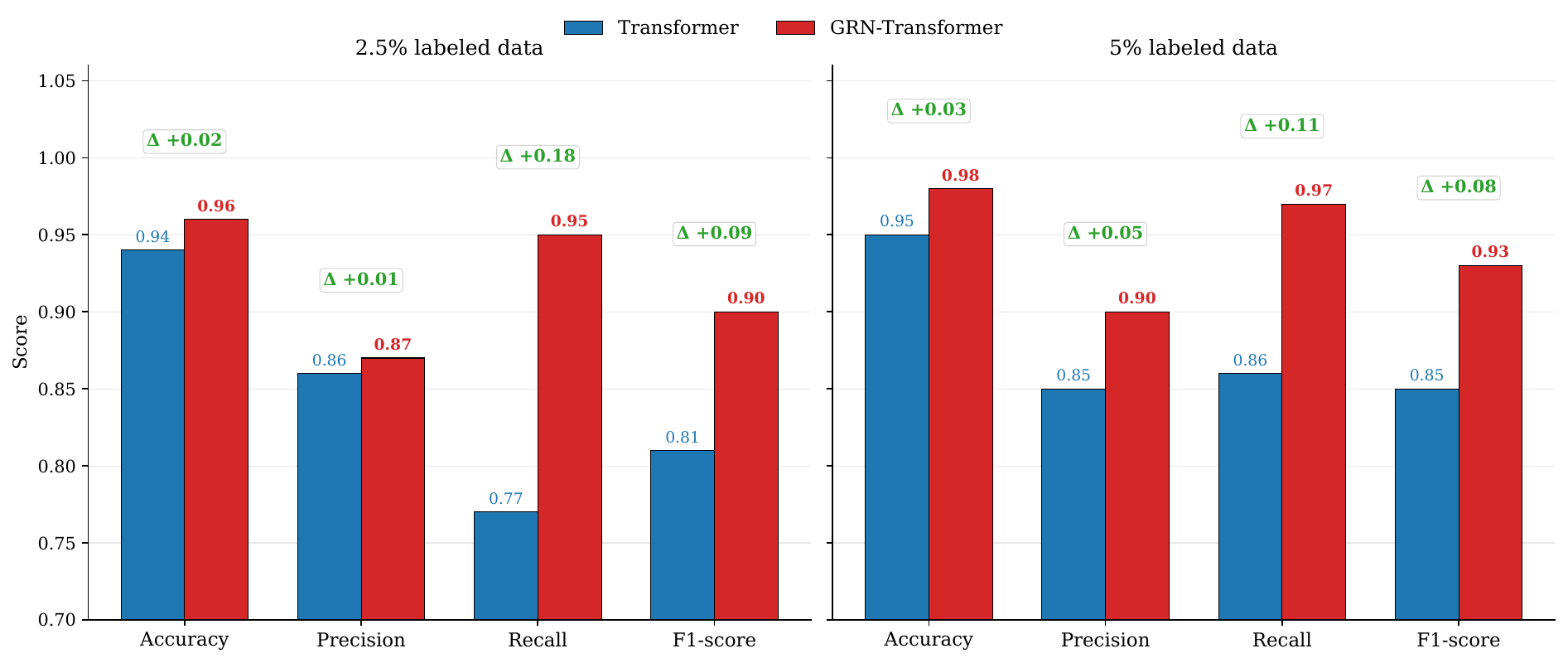} \caption{Effect of adding the Gated Residual Network (GRN) block to the Transformer classifier, evaluated on the two labeled-data regimes ($2.5\%$ and $5\%$). Bars show accuracy, precision, recall, and F1-score on the held-out test set; numeric labels above each pair report the absolute improvement $\Delta = \text{GRN-Transformer} - \text{Transformer}$. The GRN block produces a consistent improvement across all four metrics at both data regimes, with the largest gains in recall ($+0.18$ at $2.5\%$; $+0.11$ at $5\%$)---the metric most relevant for clinical artifact detection, where missed artifacts propagate into downstream physiological estimates such as SpO\textsubscript{2}. Headline F1 gains are $+0.09$ and $+0.08$ respectively.}
\label{fig:transformer-grn} 
\end{figure*}

All experiments were conducted on the PICU e-Medical infrastructure (Miircic Server) at CHUSJ, using a GPU Quadro RTX 6000 with 24~GB of memory, with code implemented in scikit-learn~\cite{scikit-learn} and Keras~\cite{chollet2015keras}. The data were split $70\%/30\%$ into training and test sets. Following~\cite{popel2018training}, the Transformer was tuned on four critical hyperparameters (model size, learning rate, batch size, and sequence length); we also applied dropout~\cite{srivastava2014dropout} ($p=0.25$), the GlorotNormal kernel initializer~\cite{glorot2010understanding}, and batch normalization~\cite{ioffe2015batch, bjorck2018understanding}. Class imbalance was addressed at training time with ADASYN oversampling~\cite{he2008adasyn}. The hyperparameters used for each classifier are summarized in Table~\ref{tab:my_hyperparameters}. 

Performance was assessed with accuracy, precision, recall (sensitivity), and F1-score~\cite{goutte2005probabilistic}: \begin{align} \text{Accuracy} &= \frac{\text{TP}+\text{TN}}{\text{TP}+\text{TN}+\text{FP}+\text{FN}} \nonumber \\ \text{Precision} &= \frac{\text{TP}}{\text{TP}+\text{FP}} \nonumber \\ \text{Recall} &= \frac{\text{TP}}{\text{TP}+\text{FN}} \nonumber \\ \text{F1} &= \frac{2 \cdot \text{Precision} \cdot \text{Recall}}{\text{Precision}+\text{Recall}} \nonumber \end{align} where TP, TN, FP, FN denote true positives, true negatives, false positives, and false negatives. 

\subsection{Overall classifier comparison} 
\label{sec:overall-comparison} 

Table~\ref{tab:my-table} compares the four neural classifiers (MLP, FCNN, BiLSTM, Transformer) with and without the GRN block, on the two labeled-data subsets that probe the small-data regime motivating this work: $2.5\%$ ($n=2{,}137$ pulses) and $5\%$ ($n=4{,}170$ pulses). The MLP, FCNN, and BiLSTM baselines gain no measurable benefit from the GRN block ($|\Delta\text{F1}| \leq 0.03$ in all cases), while the Transformer improves substantially: from F1~$=0.81$ to F1~$=0.90$ at $2.5\%$, and from F1~$=0.85$ to F1~$=0.93$ at $5\%$. The $5\%$ GRN-Transformer is the single best classifier in the table under the single-split protocol of the original benchmark, with $98\%$ accuracy and $97\%$ recall. 

The architecture-specific effect on the Transformer is shown in Fig.~\ref{fig:transformer-grn}. At $2.5\%$ labeled data, the GRN block raises accuracy from $0.94$ to $0.96$, precision from $0.86$ to $0.87$, recall from $0.77$ to $0.95$, and F1 from $0.81$ to $0.90$. At $5\%$ labeled data, the gains are $0.95 \rightarrow 0.98$ (accuracy), $0.85 \rightarrow 0.90$ (precision), $0.86 \rightarrow 0.97$ (recall), and $0.85 \rightarrow 0.93$ (F1). We note two observations. First, the largest single improvement is in recall ($+0.18$ at $2.5\%$; $+0.11$ at $5\%$), directly addressing the clinical priority of catching artifacts before they propagate into downstream SpO\textsubscript{2} estimates. Second, precision improves in parallel with recall, rather than the usual precision/recall trade-off when a model becomes more sensitive, indicating that the GRN block produces a better-calibrated decision boundary rather than merely shifting the operating threshold. We interpret the asymmetry between the Transformer and the shallower baselines as evidence that the GRN acts as a small-data regularizer whose benefit scales with model capacity: the shallower architectures already generalize adequately from limited PPG data, whereas the high-capacity Transformer, known to overfit small clinical datasets~\cite{claramacabiau2023, lee2021vision}, is precisely where a gated identity pathway provides the most leverage. 

\subsection{Interpretability of the GRN block} 
\label{sec:interpretability}

\begin{figure*}[t] 
\centering \includegraphics[width=0.95\textwidth]{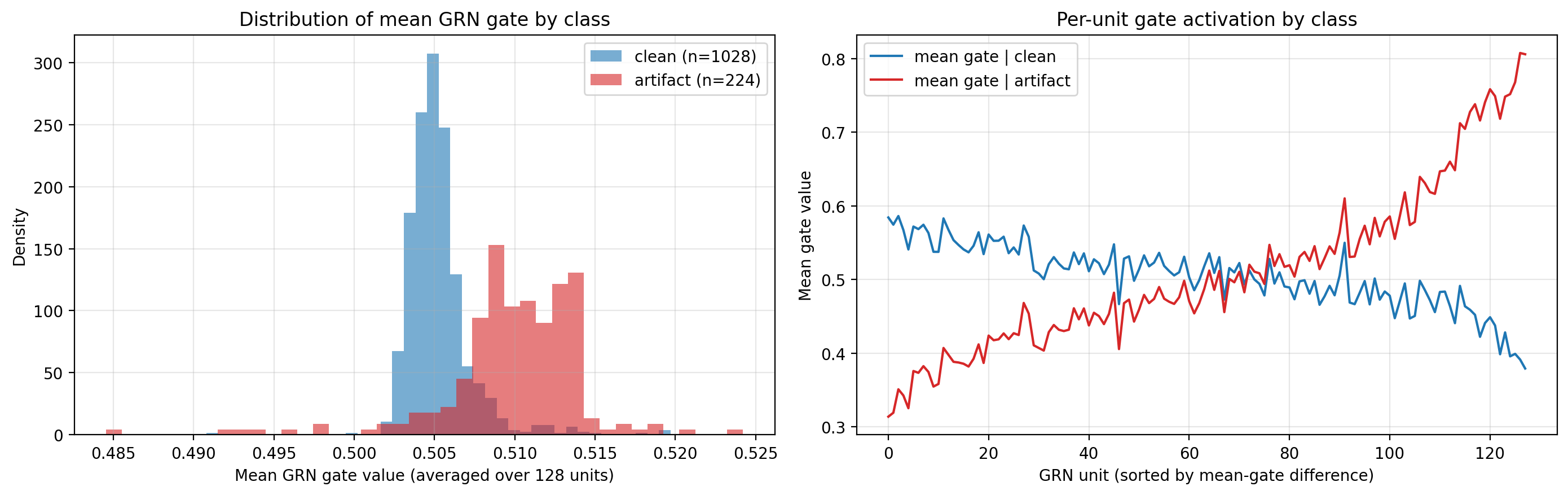} 
\caption{Interpretability analysis of the GRN block in the trained GRN-Transformer. The GRN's gated-linear-unit (GLU) produces a $128$-dimensional sigmoid gate vector per input pulse; we examine these gate values on the held-out test set ($n=1{,}252$ pulses; $1{,}028$ clean and $224$ artifact). \emph{Left:} distribution of the per-pulse mean gate value (averaged over the $128$ units), separately for clean (blue) and artifact (red) pulses. The two distributions are tightly clustered but visibly shifted: clean pulses concentrate around $0.505$ while artifact pulses concentrate around $0.510$, with extended right-tail mass up to $\sim 0.525$. The shift is statistically significant (Mann--Whitney $U$ test, $p \approx 10^{-74}$; Kolmogorov--Smirnov statistic $0.77$, $p \approx 10^{-110}$), although the absolute difference between class means is small ($\Delta = 0.005$). \emph{Right:} per-unit mean gate activation for the two classes, with units sorted by the mean-gate difference (artifact minus clean). The two curves diverge monotonically: a subset of GRN units consistently opens more for artifact pulses while a complementary subset opens more for clean pulses, indicating that the GRN has learned a class-conditional gating pattern rather than a uniform global response.} 
\label{fig:grn-gates} 
\end{figure*}

A natural concern with any architectural addition is whether the new parameters are functionally engaged or simply provide additional capacity that the optimizer never specializes. To address this, we inspected the GRN block's GLU activations on the held-out test set from the $5\%$-labeled subset. Each pulse produces a $128$-dimensional sigmoid gate vector at the GLU output, which we treated as the hypothesized mechanism by which the GRN modulates the Transformer's penultimate representation; if the block is functionally active, gate activations should differ systematically between clean and artifact pulses. 


Fig.~\ref{fig:grn-gates} reports this analysis. The per-pulse mean gate value (left panel) differs between classes by a small but statistically significant margin. We compared the two empirical distributions with the two-sample Kolmogorov--Smirnov test~\cite{smirnov1948table}, which measures the maximum vertical distance between the cumulative distribution functions, and the Mann--Whitney $U$ test~\cite{mann1947test},  which tests whether one distribution is stochastically larger than the other. The KS statistic is $D = 0.77$ ($p \approx 10^{-110}$) and the Mann--Whitney $p$-value is $\approx 10^{-74}$, despite the absolute mean shift being only $\Delta = 0.005$ (clean mean $\approx 0.505$; artifact mean $\approx 0.510$). Both tests were computed using \texttt{scipy.stats.ks\_2samp} and \texttt{scipy.stats.mannwhitneyu}~\cite{virtanen2020scipy}.

The high statistical significance combined with the small effect size indicates that the gating is consistent rather than dramatic; the GRN does not strongly suppress or amplify any single feature, but does shift its gating posture systematically as a function of input class. The per-unit analysis (right panel) clarifies the mechanism: when the $128$ GRN units are sorted by the per-class mean-gate difference, the two curves separate monotonically over the entire range, with subsets of units specializing in clean and artifact pulses, respectively.  While we do not claim the gating pattern is the sole explanation for the GRN-Transformer's improved performance, it provides mechanistic evidence that the block is functionally engaged in the classification task. This is consistent with the theoretical motivation of the gated identity pathway~\cite{savarese2017residual, dauphin2017language}, in which the GLU is designed to allow the network to learn when, and to what degree, the GRN's nonlinear transformation should contribute to the output. A complementary information-theoretic characterization of this gating behavior, using mutual-information neural estimation, is developed in our follow-up work~\cite{le2026transformer}.


\subsection{Computational complexity} 
\label{sec:complexity}

\begin{table*}[!ht]
\centering
\caption{Complexity Comparison Between Transformer and GRN-Transformer.}
\begin{tabular}{|l|c|c|}
\hline
Metric & Transformer & GRN-Transformer \\ \hline
Total parameters & 61,769 & 128,073 \\ \hline
Trainable parameters & 61,769 & 128,073 \\ \hline
Layer structure & Standard Transformer + MLP & Transformer + MLP + GRN \\ \hline
Computation type added by GRN & N/A & Gated Linear Operations (Dense + Sigmoid) \\ \hline
FLOPs (million) & 1,091.68  & 1,091.82  \\ \hline
MACs (million) & 545.84  & 545.91  \\ \hline
Training time (s) & 355.5  & 376.5  \\ \hline
Total inference time per test fold (s) & 1 & 1  \\ \hline
\end{tabular}
\label{tab:complexity_comparison}
\end{table*}

Table~\ref{tab:complexity_comparison} compares the computational profile of the Transformer and GRN-Transformer. Adding the GRN block roughly doubles the parameter count, from $61{,}769$ to $128{,}073$, due to the additional dense layers, gating units, and layer normalization. The computational impact, however, is negligible: FLOPs ($1{,}091.68$ vs $1{,}091.82$~M) and MACs ($545.84$ vs $545.91$~M) are nearly identical, because the GRN contributes element-wise rather than matrix-multiplication-dominated operations. Training time increases marginally ($355.5$~s to $376.5$~s) and inference time is unchanged at $1$~s per fold. The static footprint of the trained model is $0.51$~MB of float-32 weights ($1.42$~MB on disk), making the GRN-Transformer suitable for deployment on hospital servers with constrained computational resources; the runtime characterization in Section~\ref{sec:deployment} confirms this on a CPU-only consumer laptop. 

\subsection{Comparison with semi-supervised and KNN baselines} 
\label{sec:lp-knn}

\begin{table*}[!htp]
\centering
\footnotesize
\caption {Comparison of GRN-Transformer with semi-supervised label propagation (LP) and KNN, on the 2.5\% and 5\% labeled subsets. Best result per column in bold.}
\label{tab:my-table_GRN_KNN_LP}
\begin{tabular}{lccccccccc}
\toprule
& \multicolumn{4}{c}{2.5\% Data} & & \multicolumn{4}{c}{5\% Data} \\
\cmidrule(lr){2-5} \cmidrule(lr){7-10}
Model & Acc & Pre & Rec & F1 & & Acc & Pre & Rec & F1 \\
\midrule
LP              & 0.96 & 0.87 & 0.93          & \textbf{0.90} & & 0.97          & 0.89          & 0.93          & 0.91          \\
KNN             & 0.96 & 0.87 & 0.92          & 0.89          & & 0.97          & 0.89          & 0.95          & 0.92          \\
GRN-Transformer & 0.96 & 0.87 & \textbf{0.95} & \textbf{0.90} & & \textbf{0.98} & \textbf{0.90} & \textbf{0.97} & \textbf{0.93} \\
\bottomrule
\end{tabular}
\end{table*}

\begin{figure*}[t] 
\centering 
\includegraphics[width=0.92\textwidth]{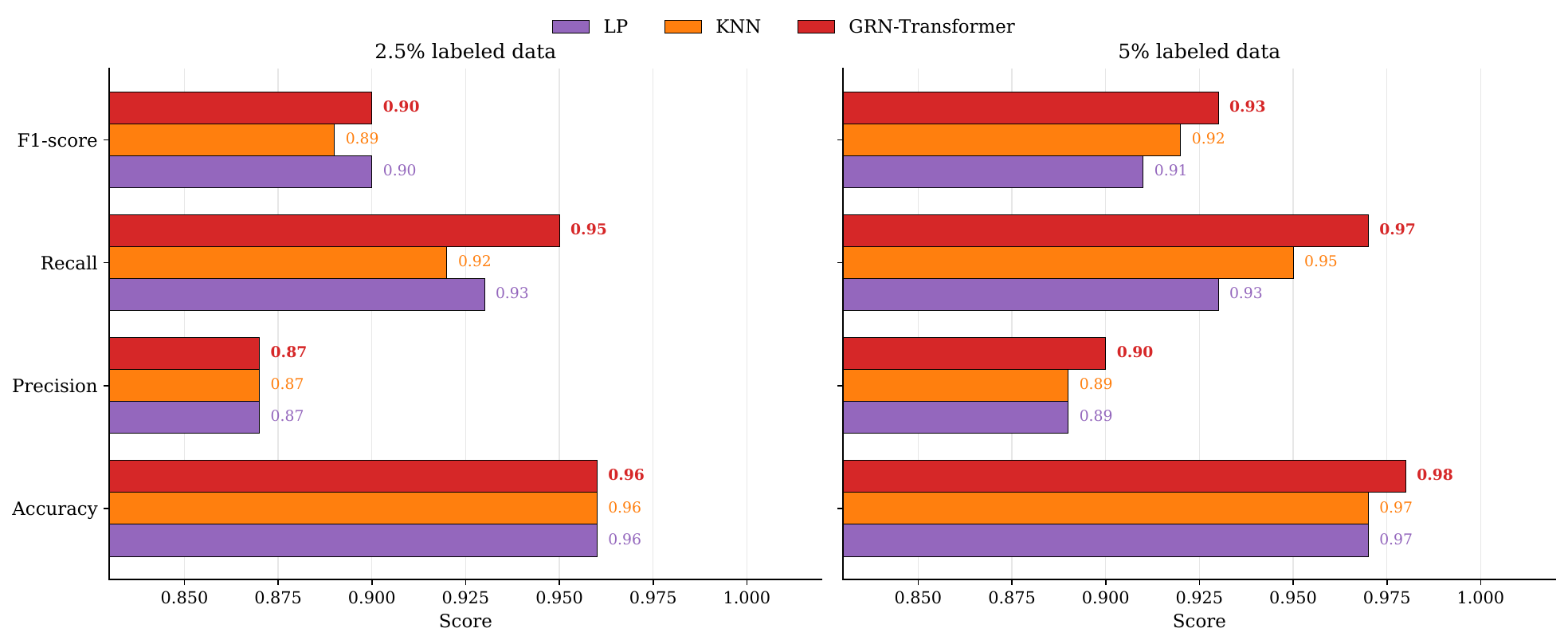} 
\caption{Comparison of the GRN-Transformer with two strong baselines from our prior work~\cite{claramacabiau2023}: semi-supervised label propagation (LP, purple) and a K-nearest-neighbor classifier (KNN, orange), at the $2.5\%$ and $5\%$ labeled-data subsets. All three methods achieve similar accuracy and precision; the GRN-Transformer's advantage lies primarily in recall (the metric most relevant to clinical artifact detection) and, consequently, in F1-score on the larger $5\%$ subset.} 
\label{fig:lp-knn-grn} 
\end{figure*}

\begin{figure*}[!htp]
	\centering
	\includegraphics[scale=0.5]{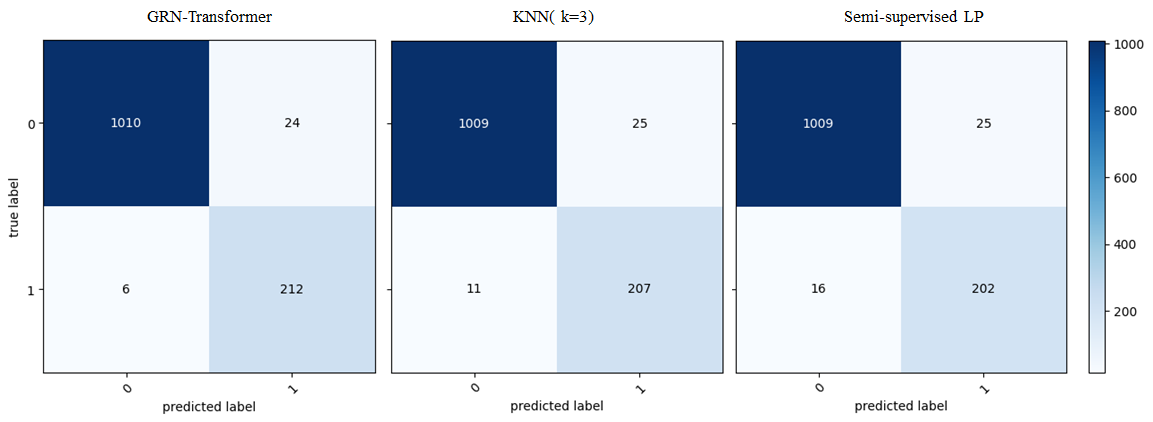}
    \caption{Confusion matrices for the GRN-Transformer, $ K$-nearest-neighbor ($K=3$), and semi-supervised label propagation (LP) classifiers on the held-out $5\%$-labeled test set ($n=1{,}252$ pulses; $1{,}034$ clean, $218$ artifact). All three methods agree closely on clean pulses ($1{,}009$--$1{,}010$ true negatives, $24$--$25$ false positives), and their disagreement is concentrated in the false-negative cell---the clinically critical category of missed artifacts. The GRN-Transformer misses $6$ artifacts (FN rate $2.8\%$), compared with $11$ for KNN ($5.0\%$) and $16$ for LP ($7.3\%$). Total misclassifications are $30$ (GRN-Transformer), $36$ (KNN), and $41$ (LP); the GRN-Transformer's advantage is driven almost entirely by reduced false negatives, consistent with the recall pattern reported in Table~\ref{tab:my-table_GRN_KNN_LP}.}
	\label{fig:confusionmatrix}
\end{figure*}

Table~\ref{tab:my-table_GRN_KNN_LP} and Fig.~\ref{fig:lp-knn-grn} compare the GRN-Transformer with the two best-performing classifiers from our prior PPG artifact-detection study~\cite{claramacabiau2023}: semi-supervised label propagation (LP) and a K-nearest-neighbor (KNN) classifier. All three methods achieve essentially the same accuracy ($0.96$--$0.98$) and precision ($0.87$--$0.90$); the GRN-Transformer's advantage lies in recall. At $2.5\%$ labeled data it reaches $0.95$ versus $0.93$ (LP) and $0.92$ (KNN); at $5\%$ the gap widens to $0.97$ versus $0.93$ and $0.95$. The corresponding F1 scores tie at $0.90$ on the $2.5\%$ subset and split $0.93/0.92/0.91$ on the $5\%$ subset in favor of the GRN-Transformer. Because recall directly determines how many true artifacts are filtered before they propagate into downstream SpO\textsubscript{2} estimates, this consistent recall advantage is the clinically meaningful distinction between the three methods, even where the absolute F1 gap appears modest. The confusion matrices on the $5\%$ test set (Fig.~\ref{fig:confusionmatrix}) show the same pattern in terms of misclassification counts: the GRN-Transformer misclassifies $30$ cases, compared with $36$ for KNN and $41$ for LP, a $16.7\%$ and $26.8\%$ relative reduction in misclassification, respectively. 

\subsection{Training-time convergence behavior}
\label{sec:convergence}

\begin{figure*}[!htp]
	\centering
	\includegraphics[scale=0.45]{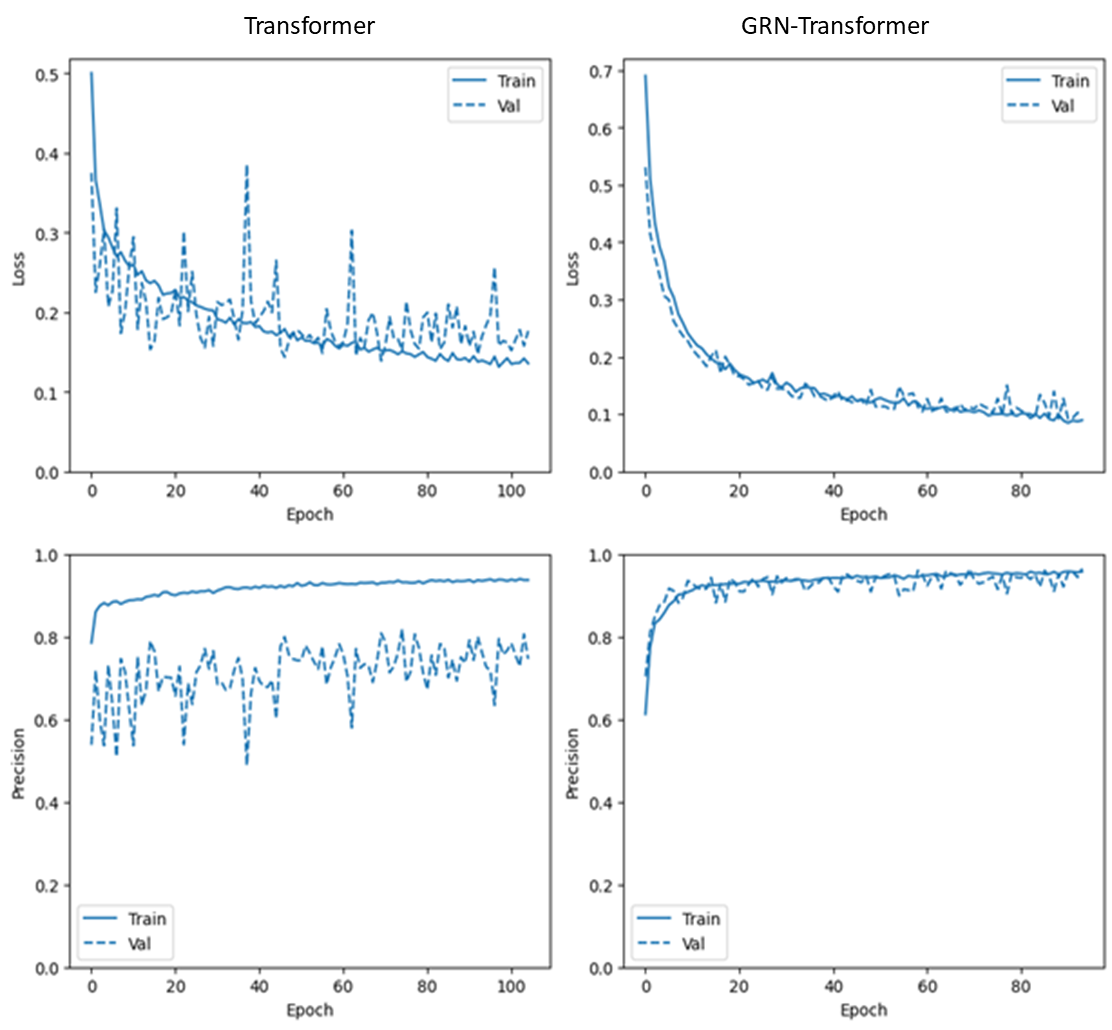}
	  \caption{Training dynamics of the standard Transformer (left column) and the GRN-Transformer (right column) on the $5\%$-labeled subset. Solid lines are training, and dashed lines are validation. \emph{Top row:} loss curves. The standard Transformer's validation loss fluctuates substantially throughout training, with repeated spikes that persist beyond epoch~80, whereas the GRN-Transformer's training and validation losses descend together smoothly and reach a tighter final value. \emph{Bottom row:} precision curves. The standard Transformer exhibits a large, sustained train--validation gap ($\sim$$0.15$--$0.25$ throughout) with highly oscillatory validation behavior, whereas the GRN-Transformer's training and validation precision rise together and remain tightly coupled. The GRN-Transformer also converges in fewer epochs ($\sim$93 vs $\sim$105 for the standard Transformer), indicating more predictable optimization on small clinical datasets.}
	\label{fig:trans_learning_curve}
\end{figure*}

Fig.~\ref{fig:trans_learning_curve} compares the training and validation curves of the Transformer and GRN-Transformer on the $5\%$-labeled subset. The standard Transformer exhibits substantial fluctuation in both loss and precision across epochs, consistent with the optimization difficulty reported for Transformers on small datasets~\cite{savarese2017residual, le2026transformer}. With the GRN block, the loss curve becomes notably smoother, and convergence is reached in fewer epochs. As the cross-validated comparison in Section~\ref{sec:crossval} shows, this stability does not by itself translate into a statistically significant clean-data accuracy advantage; the convergence behavior is, however, useful in its own right, since faster and more predictable training reduces the wall-clock cost of hyperparameter search and the retraining cycles required as additional clinical data becomes available.

\subsection{Cross-validated performance and per-fold variability} 
\label{sec:crossval} 
\begin{figure*}[t] 
\centering 
\includegraphics[width=0.95\textwidth]{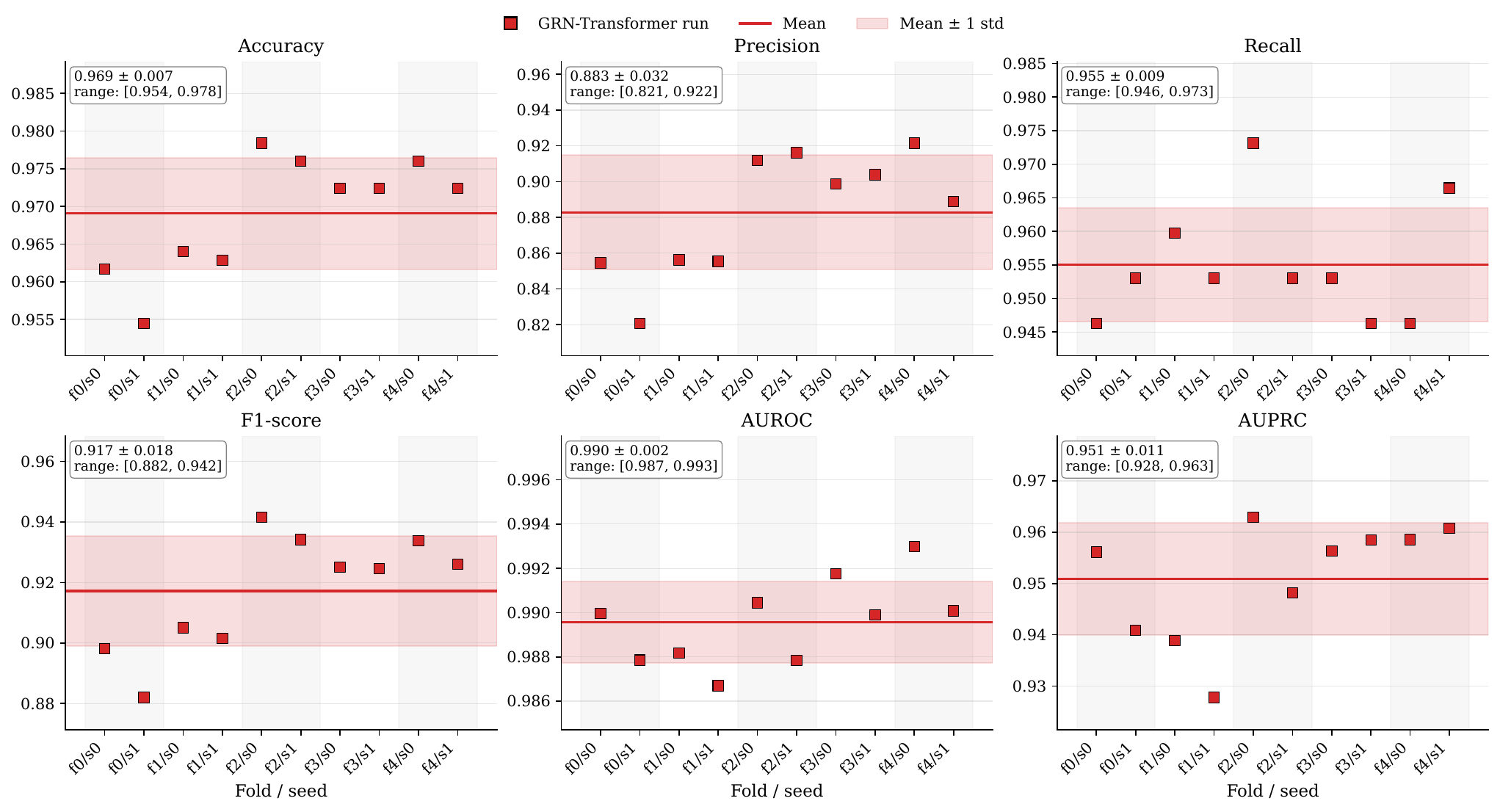} 
\caption{Cross-validated performance of the GRN-Transformer on the $5\%$-labeled dataset, evaluated under stratified $5$-fold cross-validation with two random seeds per fold ($n=10$ runs). Each red square is one (fold, seed) run. The solid red line is the mean across runs and the shaded band is the mean $\pm 1$ standard deviation. The alternating grey background bands group runs by fold so that fold-to-fold variability can be distinguished from seed-to-seed variability. The annotation box in each panel reports the mean, standard deviation, and full range. Across the $10$ runs the model is consistent on accuracy ($0.969 \pm 0.007$), recall ($0.955 \pm 0.009$), AUROC ($0.990 \pm 0.002$), and AUPRC ($0.951 \pm 0.011$), but exhibits larger per-run variability in precision ($0.883 \pm 0.032$, range $[0.821, 0.922]$) and consequently in F1-score ($0.917 \pm 0.018$).}
\label{fig:grn-cv-limitations} 
\end{figure*}

The single-split evaluations of Tables~\ref{tab:my-table} and~\ref{tab:my-table_GRN_KNN_LP} achieve $0.98$ accuracy and $0.93$ F1 at $5\%$ labeled data, but single-split numbers can overstate or understate true performance depending on which patients end up in the test set. To estimate the run-to-run variability that a clinical deployment would actually face, we re-evaluated the GRN-Transformer on the $5\%$-labeled dataset using stratified $5$-fold cross-validation (CV) with two random seeds per fold, yielding $n=10$ independent training and evaluation runs.

Fig.~\ref{fig:grn-cv-limitations} reports per-(fold, seed) results. The model is consistent on most metrics: accuracy within a $2.4$-point band ($0.954$--$0.978$), recall within $2.7$ points ($0.946$--$0.973$), AUROC within $0.7$ points ($0.987$--$0.993$), and AUPRC within $3.5$ points ($0.928$--$0.963$). Precision varies more across runs ($0.883 \pm 0.032$, range $[0.821, 0.922]$), and this dispersion propagates into F1 ($0.917 \pm 0.018$, range $[0.882, 0.942]$). The fact that precision varies more than recall indicates that fold-to-fold changes affect the model's false-positive behavior more than its ability to detect true artifacts, a clinically useful property, since false alarms are easier to tolerate than missed ones, but one whose stability under broader patient-population shifts remains to be confirmed. The mean CV F1-score of $0.917$ is approximately one percentage point lower than the single-split F1 of $0.93$ reported earlier, indicating that the single-split result was modestly favorable rather than typical. All $10$ runs were drawn from the same CHUSJ PICU cohort; the variability we observe, therefore, reflects only within-site sampling effects and is likely an \emph{underestimate} of the variability the model would exhibit when evaluated on independent cohorts from other pediatric ICUs. External validation at least one additional PICU site and a multi-annotator agreement study will be necessary before any claim of clinical deployment readiness can be made. 

\subsection{Robustness under realistic PPG signal degradation} 
\label{sec:robustness}

\begin{figure*}[t] 
\centering 
\includegraphics[width=\textwidth]{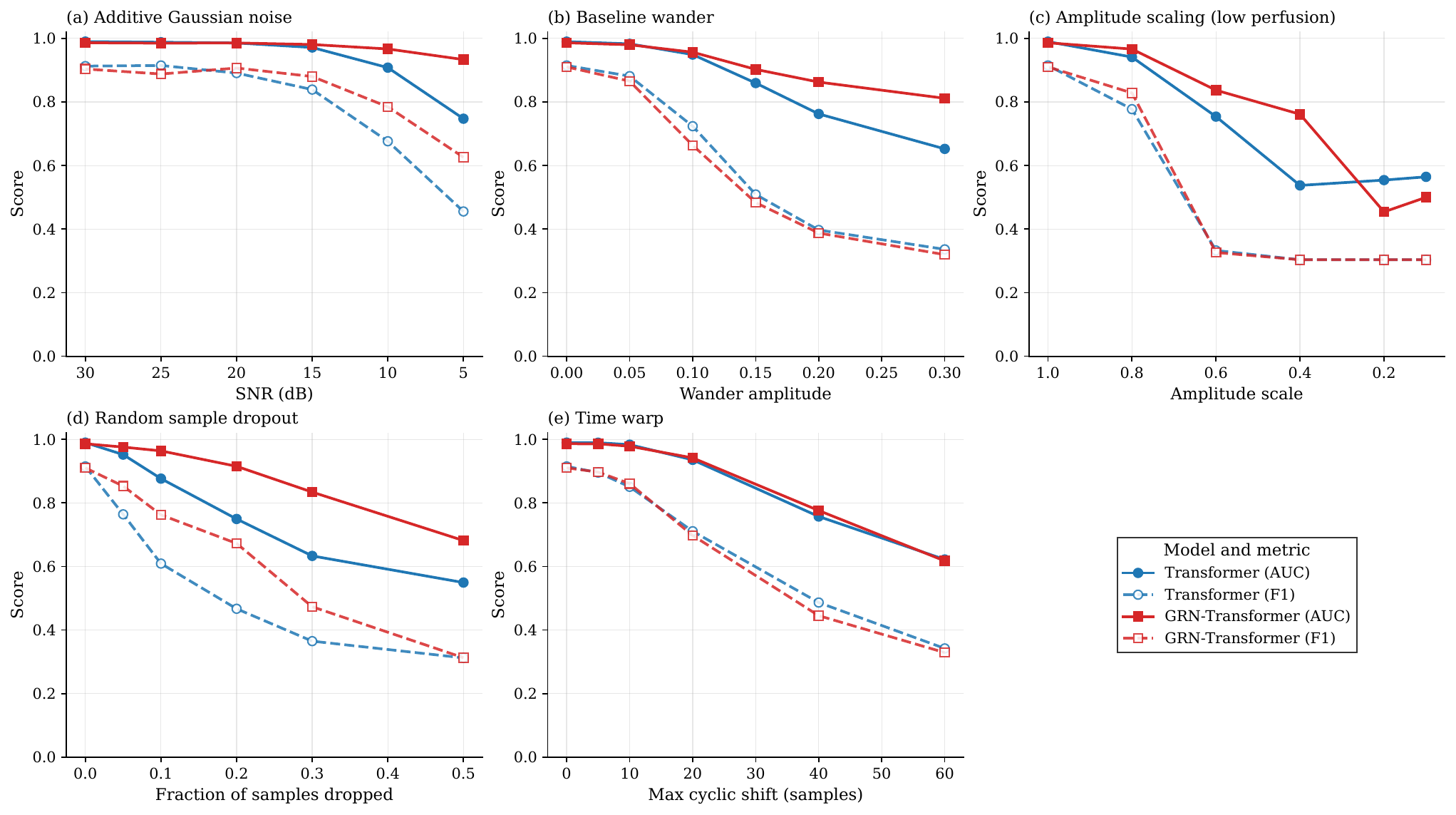} 
\caption{Robustness of the Transformer (blue) and GRN-Transformer (red) under five controlled signal degradations applied to the held-out test set of fold~0. Solid lines and filled markers denote AUROC; dashed lines and open markers denote F1 at the F1-optimal threshold calibrated on the clean test set. Both models start from the same clean-data performance (AUROC~$\approx 0.99$, F1~$\approx 0.91$). (a)~Additive Gaussian noise at decreasing signal-to-noise ratios. (b)~Low-frequency baseline wander superimposed on each pulse, simulating slow motion and respiratory artifacts. (c)~Multiplicative amplitude scaling, a proxy for low peripheral perfusion. (d)~Random per-sample dropout, simulating intermittent sensor disconnection. (e)~Random cyclic time shift, simulating heart-rate variability and beat-detection misalignment. The $x$-axis is oriented so that severity increases from left to right in every panel. The GRN-Transformer degrades substantially more gracefully than the standard Transformer under noise (a), baseline wander (b), and sample dropout (d), the three perturbations most representative of artifacts encountered in clinical PPG monitoring.} 
\label{fig:robustness} 
\end{figure*}

Cross-validated clean-data parity does not by itself justify deployment in a PICU, where signal quality varies continuously with patient movement, sensor positioning, and peripheral perfusion. We therefore evaluated the Transformer and GRN-Transformer under five controlled perturbations applied to the held-out test set, each chosen to mimic a distinct source of clinical signal degradation documented in the PPG-noise literature~\cite{park2022photoplethysmogram}. For each perturbation, both models were trained once on clean data and evaluated on test sets perturbed at six increasing levels of severity (Fig.~\ref{fig:robustness}). The five perturbations are: 

\begin{itemize} 

\item \textbf{Additive Gaussian noise} $\mathcal{N}(0, \sigma^{2})$ at SNR levels from $30$ to $5$~dB: broadband electronic and sensor noise, the standard baseline for PPG signal-quality assessment~\cite{dao2016robust}. 
\item \textbf{Low-frequency baseline wander} (sinusoidal, $0.05$--$0.5$~Hz, amplitude up to $0.3$ of the pulse range): respiratory motion and slow body movement at the sensor-skin interface~\cite{park2022photoplethysmogram}. 
\item \textbf{Multiplicative amplitude scaling} ($\alpha \in \{1.0,0.8,0.6,0.4,0.2\}$):  a controlled proxy for low peripheral perfusion, particularly relevant in the PICU where vasoconstriction, hypotension, or hypothermia commonly reduce PPG amplitude~\cite{giuliano2023comparative}. 
\item \textbf{Random per-sample dropout} (fraction $p \in \{0,0.1,\ldots,0.5\}$):  intermittent optical contact between the sensor and skin during patient movement, a class of corruption that conventional denoising methods rarely address ~\cite{park2022photoplethysmogram}.
\item \textbf{Random cyclic time shifts} ($k \in \{0,10,\ldots,60\}$ samples): segmentation misalignment from heart-rate variability and beat-detection errors upstream of the classifier~\cite{elgendi2012analysis}. 

\end{itemize}

On the three perturbations most representative of clinically encountered artifacts, noise, baseline wander, and sample dropout (Fig.~\ref{fig:robustness}a,~b,~d), the GRN-Transformer degrades substantially more gracefully than the standard Transformer. At a $5$~dB signal-to-noise ratio, the GRN-Transformer retains AUROC $0.93$ compared with $0.75$ for the Transformer ($+18$ points), and F1 $0.63$ versus $0.46$ ($+17$ points). Under baseline amplitude wander of $0.30$, the GRN-Transformer maintains AUROC $\approx 0.81$ versus $\approx 0.65$ ($+16$ points). Under $50\%$ random sample dropout, AUROC is $\approx 0.68$ versus $\approx 0.55$ ($+13$ points). The advantage is visible across the full severity range, not only at the worst levels, indicating that the effect is a property of the regularized representation rather than a tail-of-distribution artifact. 

Under extreme amplitude scaling (Fig.~\ref{fig:robustness}c), both models collapse below AUROC $0.6$ once the signal is reduced to $20\%$ of original amplitude; the GRN-Transformer retains a small margin in the moderate-degradation regime ($\geq 0.6$, scale) but neither model remains clinically useful at the extreme. Under random time-warping (Fig.~\ref{fig:robustness}e), the two models degrade in lockstep, indicating that neither has learned representations invariant to gross beat-detection misalignment; this class of failure would need to be addressed at the segmentation stage rather than the classifier. Taken together, these results refine the interpretation of clean-data parity: the two architectures are interchangeable when input quality is high, but diverge sharply when realistic signal corruption is introduced. The GRN block, which has no measurable effect on clean-data classification accuracy, serves as a robustness regularizer that protects performance precisely in the conditions where a PPG artifact detector is most needed. 

\subsection{Deployment benchmarks} 
\label{sec:deployment}

\begin{figure*}[!htp] 
\centering 
\includegraphics[width=\textwidth]{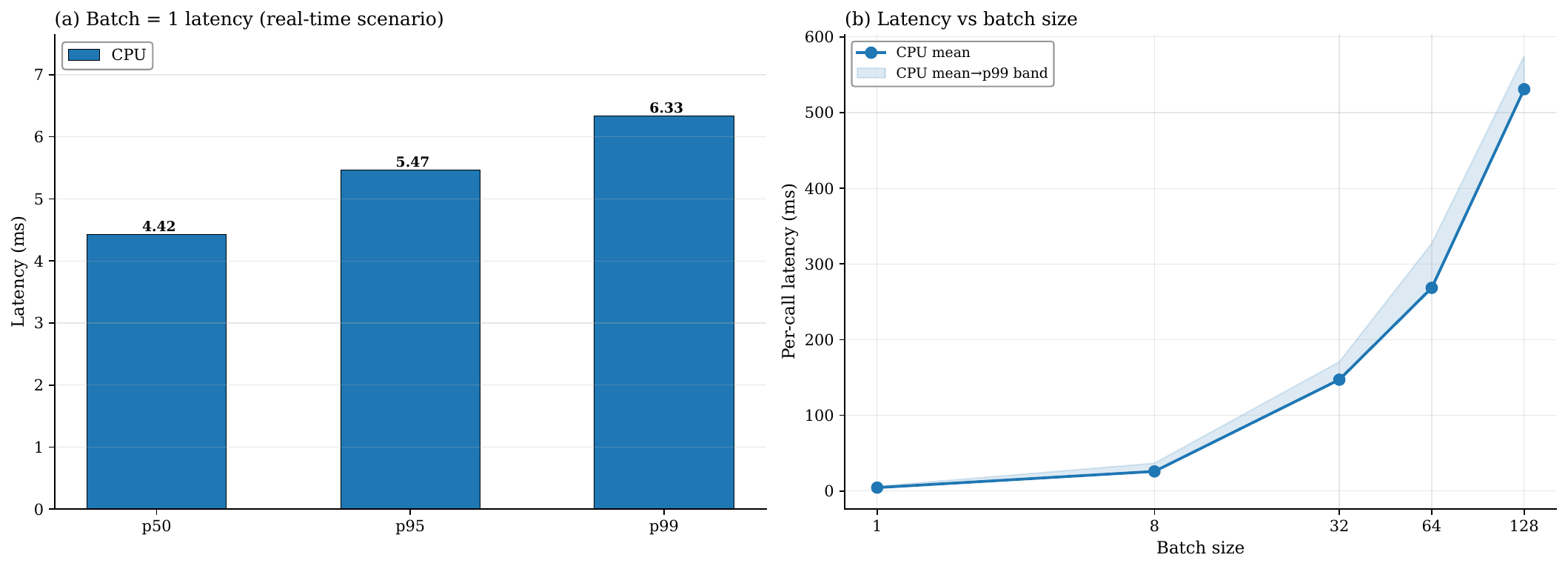} 
\caption{GRN-Transformer deployment benchmarks on a consumer laptop (AMD64, 8~physical cores, 16~GB~RAM, CPU-only inference; TensorFlow~2.16.1, Python~3.11.7), chosen as a conservative proxy for the resource-constrained workstations available on hospital networks. (a)~Per-call inference latency at batch size 1, the operating mode of a real-time clinical alarm filter. The median (p50), 95\textsuperscript{th} percentile (p95), and 99\textsuperscript{th} percentile (p99) latencies are reported over $10^{3}$ measurements after $50$ warm-up runs. The p99 latency of $6.33$~ms is over three orders of magnitude below the $30$~s duration of each PPG window, indicating ample timing headroom for real-time deployment. (b)~Per-call latency as a function of batch size (log\textsubscript{2} spacing on $x$), with the shaded band spanning the mean-to-p99 range. The near-linear scaling of latency with batch size, together with the narrow mean-to-p99 spread, demonstrates predictable throughput-mode behavior suitable for retrospective batch processing of stored PPG records.} 
\label{fig:benchmarks} 
\end{figure*}

To assess whether the GRN-Transformer can run in the compute-constrained environments typical of clinical workstations, we benchmarked it on a consumer laptop (AMD64, 8~physical cores, 16~GB~RAM, CPU-only inference; TensorFlow~2.16.1, Python~3.11.7). This choice is deliberately conservative, approximating a lower bound on the compute likely to be available on a hospital workstation. Fig.~\ref{fig:benchmarks} summarizes the results. 

In the real-time scenario (single-window inference at batch size $1$, panel~a), the model achieves a median latency of $4.42$~ms, with $95\%$ of calls completing within $5.47$~ms and $99\%$ within $6.33$~ms. Since each PPG analysis window represents $30$~s of monitoring data, a worst-case $6.33$~ms inference consumes only $0.02\%$ of the available real-time budget per window; equivalently, the model can classify $\sim$$4{,}700$ pulses for every pulse arriving from the bedside monitor. The narrow median-to-p99 spread ($\Delta = 1.91$~ms) indicates an absence of rare high-latency outliers, a property required for alarm-triggering pipelines where worst-case behavior matters as much as the average~\cite{popel2018training}. 

For retrospective batch processing (panel~b), per-call latency scales approximately linearly with batch size, from $\sim$$5$~ms at batch~$1$ to $\sim$$540$~ms at batch~$128$. The corresponding throughput at batch~$128$ is approximately $237$~pulses~s$^{-1}$, sufficient to reprocess on the order of $8 \times 10^{5}$ pulses per hour on a single CPU, more than adequate to clean a full PICU's daily PPG load offline on commodity hardware. Combined with the small static footprint reported in Section~\ref{sec:complexity}, these measurements support the claim that the GRN-Transformer can be deployed in a PICU environment without requiring a dedicated GPU or specialized inference accelerator. 

\subsection{Retrospective clinical-impact simulation} 
\label{sec:clinical-impact}

\begin{figure*}[!htp] 
\centering \includegraphics[width=0.95\textwidth]{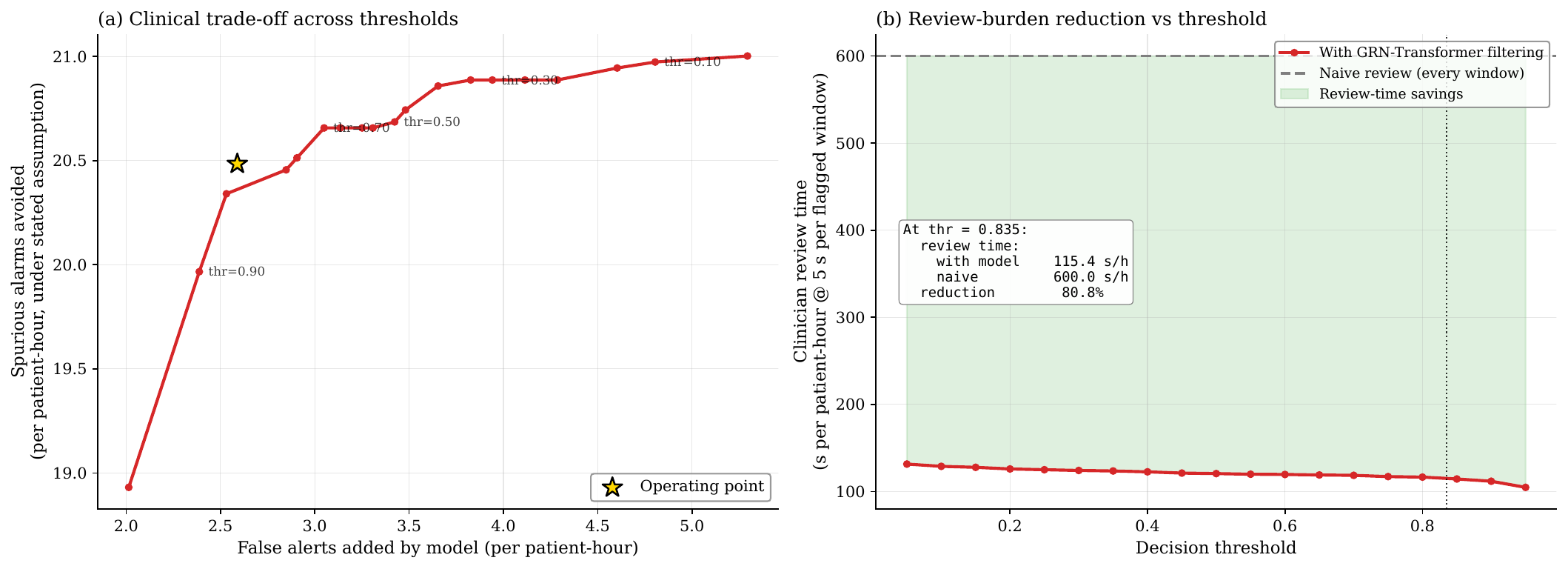} 
\caption{Retrospective impact simulation of the GRN-Transformer on the held-out test set, parameterized by the decision threshold. (a)~The clinical trade-off in deployment units: each point on the curve represents one decision threshold, with the $x$-axis showing the rate of false alerts the model would add to the clinical workflow (in patient-hour units) and the $y$-axis showing the rate of spurious alarms it would avoid, under the stated assumption that each undetected PPG artifact would otherwise trigger exactly one downstream SpO\textsubscript{2} alarm. The gold star marks the F1-optimal operating threshold ($0.835$), at which the model avoids $\sim$20.5 spurious alerts per patient-hour while incurring $\sim$2.6 false alerts per patient-hour. (b)~Clinician review-time burden as a function of decision threshold. The grey dashed line is the hypothetical naive-review baseline ($600$~s per patient-hour, corresponding to reviewing every $30$-second PPG window at $5$~s per review); the red curve is the review time required when the clinician verifies only model-flagged windows. The green-shaded band represents the implied time-saving region. At the operating threshold, model-assisted review requires $115.4$~s, an $80.8\%$ reduction relative to the naive baseline.}
\label{fig:clinical-impact} 
\end{figure*}

A concern in deploying any artifact-detection model is whether the filter would meaningfully reduce false-alarm burden without introducing unacceptable false-alert overhead. To estimate the magnitudes involved, we conducted a retrospective simulation on the GRN-Transformer's held-out test predictions across decision thresholds spanning the full probability range. We note at the outset that the resulting estimates are \emph{retrospective and assumption-conditional}: real deployment behavior, including clinician trust, alarm-system debouncing, and population shift, would need to be verified prospectively through paired alarm logs and clinician evaluations. Fig.~\ref{fig:clinical-impact} summarises the two clinically meaningful quantities. 

First, panel~(a) translates the model's precision and recall behavior into clinical units. Under the conservative assumption that each undetected artifact would trigger one downstream alarm in the absence of filtering, the model at its F1-optimal operating threshold ($0.835$) would avoid approximately $20.5$ spurious alarms per patient-hour at a cost of $2.6$ additional false alerts per patient-hour, a $\sim$$8$:$1$ ratio at this operating point. The curve's shape (steep, then plateauing) indicates that the operating point lies at the knee of the trade-off, beyond which further threshold reductions yield diminishing returns in avoided alarms, with a sharply increasing false-alert cost. This provides a basis for selecting site-specific operating thresholds compatible with local alarm-fatigue tolerance. 

Second, panel~(b) addresses clinician review-time burden. As a working assumption in the range of expert PPG segment-labeling times reported in~\cite{li2012dynamic} ($6$~s per labeled segment), we assume $5$~s of clinician time per flagged $30$-second window. One patient-hour contains $120$ non-overlapping $30$-second PPG windows; a hypothetical policy that reviewed every window would therefore require $120 \times 5 = 600$~s of clinician time per patient-hour. With the GRN-Transformer filtering at the operating threshold, only the $\sim$$23$ windows per patient-hour that the model flags need review, requiring $\sim$$115.4$~s per patient-hour, or roughly $2$~minutes. We emphasize that the $600$-s/h figure is a hypothetical upper bound rather than a description of current PICU practice; the clinically meaningful quantities are the absolute $\sim$$2$~minutes per patient-hour of model-assisted review time, the threshold-insensitivity of this estimate (spanning $\sim$$70$--$130$~s per patient-hour across the clinically plausible range $0.3 \leq \mathrm{thr} \leq 0.9$), and the linearity of both quantities in the per-window review-time assumption (so a reader preferring a different value can rescale accordingly without affecting the qualitative findings). Prospective validation with paired alarm-system telemetry and clinician evaluation remains a necessary next step. 

\section{Limitations} 
\label{sec:limitations} 
The results reported in this study have several limitations that should temper any claim of immediate clinical readiness. First, all training and evaluation data were drawn from a single-center cohort at CHUSJ; the per-fold variability quantified in Section~\ref{sec:crossval} therefore reflects only within-site sampling effects and is almost certainly an underestimate of the variability the model would exhibit on independent PICU cohorts. Second, the clinical-impact estimates in Section~\ref{sec:clinical-impact} are retrospective simulations that rest on two stated assumptions: that each undetected artifact would trigger one downstream alarm, and that a clinician spends approximately $5$~s verifying each model-flagged window. Both assumptions are defensible as orders of magnitude, but neither has been empirically calibrated against paired bedside alarm telemetry or clinician workflow timings. Third, annotation in this study was performed by a single human expert with automated cross-checking; no inter-rater agreement study was conducted, leaving residual label-noise variability unmeasured. 

The scope of the analysis is also bounded in several methodological respects. The artifact-detection task was framed as a binary classification (artifact versus normal), a simplification that elides finer-grained categories of waveform quality, such as moderate-perfusion or partially recoverable signals. Beyond binary labeling, our comparison considered only supervised and semi-supervised learning paradigms; unsupervised approaches such as recurrent autoencoders~\cite{azar2021deep} were not evaluated and remain a reasonable direction for future work. Finally, the robustness analysis (Section~\ref{sec:robustness}) showed that neither the Transformer nor the GRN-Transformer is invariant to gross time-warping, suggesting that beat-detection misalignment, a common upstream failure mode in PPG segmentation, would need to be addressed at the pre-processing rather than the classification stage. Variants of the GLU~\cite{le2026transformer, shazeer2020glu} were also unexplored and could plausibly further improve the regularizer's behavior. 

These limitations define the validation pipeline required before the GRN-Transformer can be considered for prospective deployment: external validation on at least one additional pediatric ICU site, calibration of the clinical-impact assumptions against real alarm and review-time logs, a multi-annotator agreement study, and an extension of the robustness battery to include the segmentation-stage failures identified above. None of these is a research barrier; each is a translational engineering step that this paper does not attempt to complete.

\section{Conclusion} 
\label{sec:conclusion} 
This work introduced the \emph{GRN-Transformer}, a small-data regularization strategy for Transformer-based artifact detection in clinical PPG signals, in which a single Gated Residual Network block is applied to the encoder stack's pooled output. On the PICU PPG dataset from CHUSJ, the GRN-Transformer markedly improves recall over a standard Transformer baseline (from $0.86$ to $0.97$ at $5\%$ labeled data) while preserving precision, and the cross-validated analysis indicates this advantage is driven by improved robustness and consistency rather than by raw clean-data accuracy. Gate-level inspection confirms that the GRN block is functionally engaged: its gating units develop class-conditional activation patterns distinguishing artifact and clean pulses. Beyond classification quality, the GRN-Transformer satisfies the deployment constraints typical of a single-center pediatric ICU. It runs at $6.33$~ms 99\textsuperscript{th}-percentile latency on a CPU-only consumer laptop, occupies $1.4$~MB of disk space, and degrades gracefully under realistic PPG signal corruption (additive noise, baseline wander, intermittent sensor dropout). A retrospective clinical-impact simulation, under stated assumptions, suggests the model could meaningfully reduce artifact-induced alarm burden when used as an input filter for PPG-driven CDSS. These results support the GRN-Transformer as a deployment-ready candidate for translational use in pediatric intensive care, with the per-fold variability, robustness, and latency evidence required to support that claim. Prospective validation on independent PICU cohorts, paired alarm-system telemetry, and clinician evaluation of flagged windows remain necessary next steps before the model can be adopted in routine clinical workflows.



\bibliographystyle{IEEEtran}
\bibliography{IEEEabrv,Bibliography}
\end{document}